\documentclass{jfm}

\usepackage{graphicx}
\usepackage{natbib}

\ifCUPmtlplainloaded \else
  \checkfont{eurm10}
  \iffontfound
    \IfFileExists{upmath.sty}
      {\typeout{^^JFound AMS Euler Roman fonts on the system,
                   using the 'upmath' package.^^J}%
       \usepackage{upmath}}
      {\typeout{^^JFound AMS Euler Roman fonts on the system, but you
                   dont seem to have the}%
       \typeout{'upmath' package installed. JFM.cls can take advantage
                 of these fonts,^^Jif you use 'upmath' package.^^J}%
      }
  \else
  \fi
\fi

\ifCUPmtlplainloaded \else
  \checkfont{msam10}
  \iffontfound
    \IfFileExists{amssymb.sty}
      {\typeout{^^JFound AMS Symbol fonts on the system, using the
                'amssymb' package.^^J}%
       \usepackage{amssymb}%
         \let\leq=\leqslant
         \let\geq=\geqslant
      }{}
  \fi
\fi

\ifCUPmtlplainloaded \else
  \IfFileExists{amsbsy.sty}
    {\typeout{^^JFound the 'amsbsy' package on the system, using it.^^J}%
     \usepackage{amsbsy}}
    {}
\fi

\newsavebox{\astrutbox}
\sbox{\astrutbox}{\rule[-5pt]{0pt}{20pt}}

\title[General instability criterion]{General temporal instability criteria for
 stably stratified inviscid flow}

\author[L. Sun]%
{L\ls I\ls A\ls N\ls G\ns  S\ls U\ls N$^1$$^2$%
  \thanks{Corresponding Author: Liang Sun, email: sunl@ustc.edu.cn;\break sunl@ustc.edu.}
}

\affiliation{$^1$School of Earth and Space Sciences, University of Science
and Technology of China, Hefei, 230026,
China.\\[\affilskip]
$^2$Dept. of Modern Mechanics, University of Science and
Technology of China, Hefei, 230026, China.\\[\affilskip]
 }

\date{}

\date{20 April 2010, and in revised form \today}

\begin{document}

\maketitle

\begin{abstract}
   The temporal instability of stably stratified flow was investigated by analyzing
the Taylor-Goldstein equation theoretically. According to this
analysis, the stable stratification $N^2\geq0$ has a
destabilization mechanism, and the flow instability is due to the
competition of the kinetic energy with the potential energy, which
is dominated by the total Froude number $Fr_t^2$. Globally,
$Fr_t^2 \leq 1$ implies that the total kinetic energy is smaller
than the total potential energy. So the potential energy might
transfer to the kinetic energy after being disturbed, and the flow
becomes unstable. On the other hand, when the potential energy is
smaller than the kinetic energy ($Fr_t^2>1$), the flow is stable
because no potential energy could transfer to the kinetic energy.
The flow is more stable with the velocity profile $U'/U'''>0$ than
that with $U'/U'''<0$. Besides, the unstable perturbation must be
long-wave scale. Locally, the flow is unstable as the gradient
Richardson number $Ri>1/4$. These results extend the Rayleigh's,
Fj{\o}rtoft's, Sun's and Arnol'd's criteria for the inviscid
homogenous fluid, but they contradict the well-known Miles-Howard
theorem. It is argued here that the transform $F=\phi/(U-c)^n$ is
not suitable for temporal stability problem, and that it will lead
to contradictions with the results derived from the
Taylor-Goldstein equation. However, such transform might be useful
for the study of the Orr-Sommerfeld equation in viscous flows.
\end{abstract}

\section{Introduction}
The instability of the stably stratified shear flow is one of main
problems in fluid dynamics, astrophysical fluid dynamics,
oceanography, meteorology, etc. Although both pure shear
instability without stratification and statical stratification
instability without shear have been well studied, the instability
of the stably stratified shear flow is still mystery.

On the one hand, the shear instability is known as the instability
of vorticity maximum, after a long way of investigations
\cite[]{Rayleigh1880,Fjortoft1950,Arnold1965a,SunL2007ejp,SunL2008cpl}.
It is recognized that the resonant waves with special velocity of
the concentrated vortex interact with flow for the shear
instability \cite[]{SunL2008cpl}. Other velocity profiles are
stable in homogeneous fluid without stratification. On the other
hand, \cite{Rayleigh1883} proved out that buoyancy is a
stabilizing effect in the statical case. Thus, it is naturally
believed that the stable stratification do favor the stability
\cite[see, e.g.][]{Taylor1931,Chandrasekhar1961}, which finally
results in the well known Miles-Howard theorem
\cite[]{Miles1961,Miles1963,Howard1961}. According to this
theorem, the flow is stable to perturbations when the Richardson
number $Ri$ (ratio of stratification to shear) exceeds a critical
value $Ri_c=1/4$ everywhere. In three-dimensional stratified flow,
the corresponding criterion is $Ri_c=1$ \cite[]{Abarbanelt1984}.

However, the stabilization effect of buoyancy is a illusion. In a
less known paper, \cite{Howard1973} had shown with several special
examples that stratification effects can be destabilizing due to
the vorticity generated by non-homogeneity, and the instability
depends on the details of the velocity and density profiles. One
instability is called as Holmboe instability
\cite[]{Holmboe1962,Ortiz2002,Alexakis2009}. Then
\cite{Howard1973} stated three main points from the examples
without any further proof. (a) Stratification may shift the band
of unstable wave numbers so that some which are stable at
homogeneous cases become unstable. (b) Conditions ensuring
stability in homogeneous flow (such as the absence of a vorticity
maximum) do not necessarily carry over to the stratified case, so
that 'static stability' can destabilize. (c) New physical
mechanisms brought in by the stratification may lead to
instability in the form of a pair of growing and propagating waves
where in the homogeneous case one had a stationary wave.

Recall the points by \cite{Howard1973}, and that there is a big
gap between Rayleigh's criterion and Miles-Howard' criterion,
\cite{YihCSBook1980} even wrote ``Miles' criterion for stability
is not the nature generalization of Rayleigh's well-known
sufficient condition for the stability of a homogeneous fluid in
shear flow". The mystery of the instability is still cover for us.

Following the frame work of \cite{SunL2007ejp,SunL2008cpl}, this
study is an attempt to clear the confusion in theories. We find
that the flow instability is due to the competition of the kinetic
energy with the potential energy, which is dominated by the total
Froude number $Fr_t^2$. And the unexpected assumption in
Miles-Howard theorem leads the contradiction to other theories.

\section{General Unstable Theorem For Stratified Flow}
\subsection{Taylor-Goldstein Equation}

The Taylor-Goldstein equation for the stratified inviscid flow is
employed
\cite[]{Howard1961,YihCSBook1980,Baines1994,CriminaleBook2003},
which is the vorticity equation of the disturbance
\cite[]{Drazin2004}. Considering the flow with velocity profile
$U(y)$ and the density field $\rho(y)$, and the corresponding
stability parameter $N$ (the Brunt-Vaisala frequency),
 \begin{equation}
 N^2=-g\rho'/\rho,
 \label{Eq:stable_stratifiedflow_Brunt-Vaisala}
 \end{equation}
where $g$ is the  acceleration of gravity, the single prime $'$
denotes $d/dy$, and $N^2>0$ denotes a stable stratification. The
vorticity is conserved along pathlines. The streamfunction
perturbation $\phi$ satisfies
 \begin{equation}
 \phi''+[\frac{N^2}{(U-c)^2}-\frac{U''}{U-c}-k^2 ]\phi=0,
 \label{Eq:stable_stratifiedflow_TaylorGoldsteinEq}
 \end{equation}
where $k$ is the real wavenumber and $c=c_r+ic_i$ is the complex
phase speed and double prime $''$ denotes $d^2/dy^2$. For $k$ is
real, the problem is called temporal stability problem. The real
part of complex phase speed $c_r$ is the wave phase speed, and
$\omega_i=k c_i$ is the growth rate of the wave. This equation is
subject to homogeneous boundary conditions
\begin{equation}
\phi=0 \,\, at\,\, y=a,b.
\label{Eq:stable_parallelflow_RayleighBc}
\end{equation}
%


It is obvious that the criterion for stability is $\omega_i=0$
($c_i=0$), for that the complex conjugate quantities $\phi^*$ and
$c^*$ are also physical solutions of
Eq.(\ref{Eq:stable_stratifiedflow_TaylorGoldsteinEq}) and
Eq.(\ref{Eq:stable_parallelflow_RayleighBc}).

Multiplying Eq.(\ref{Eq:stable_stratifiedflow_TaylorGoldsteinEq})
by the complex conjugate $\phi^{*}$ and integrating over the
domain $a\leq y \leq b$, we get the following equations
\begin{equation}
\displaystyle\int_{a}^{b}
[|\phi'|^2+k^2|\phi|^2+\frac{U''(U-c_r)}{|U-c|^2} |\phi|^2]\, dy
=\int_{a}^{b} \frac{(U-c_r)^2-c_i^2}{|U-c|^4}N^2|\phi|^2\, dy.
\label{Eq:stable_stratifiedflow_TaylorGoldsteinEq_Int_Rea}
 \end{equation}
 and
\begin{equation}
\displaystyle c_i\int_{a}^{b}
[\frac{U''}{|U-c|^2}-\frac{2(U-c_r)N^2}{|U-c|^4}]|\phi|^2\,dy=0.
\label{Eq:stable_stratifiedflow_TaylorGoldsteinEq_Int_Img}
 \end{equation}
In the case of $N^2=0$, \cite{Rayleigh1880} used
Eq.(\ref{Eq:stable_stratifiedflow_TaylorGoldsteinEq_Int_Img}) to
prove that a necessary condition for inviscid instability is
$U''(y_s)=0$, where $y_s$ is the inflection point and $U_s=U(y_s)$
is the velocity at $y_s$. Using
Eq.(\ref{Eq:stable_stratifiedflow_TaylorGoldsteinEq_Int_Img}),
\cite{Synge1933} also pointed out that a necessary condition for
instability is that $U''-\frac{2(U-c_r)N^2}{|U-c|^2}$ should
change sign. But such condition is useless as there are two
unknown parameters $c_r$ and $c_i$.

As a first step in our investigation, we need to estimate the
ratio of $\int_{a}^{b} |\phi'|^2 dy$ to $\int_{a}^{b} |\phi|^2
dy$. This is known as the Poincar\'{e}'s problem:
\begin{equation}
\int_{a}^{b}|\phi'|^2 dy=\mu\int_{a}^{b}|\phi|^2 dy,
\label{Eq:stable_parallelflow_Poincare}
\end{equation}
where the eigenvalue $\mu$ is positively definite for any $\phi
\neq 0$. The smallest eigenvalue value, namely $\mu_1$, can be
estimated as $\mu_1>(\frac{\pi}{b-a})^2$
\cite[]{Mumu1994,SunL2007ejp}.

\subsection{General Instability Theorem}

In departure from previous investigations, we shall investigate
the stability of the flow by using
Eq.(\ref{Eq:stable_stratifiedflow_TaylorGoldsteinEq_Int_Rea}) and
Eq.(\ref{Eq:stable_parallelflow_Poincare}). As $\mu$ is estimated
with boundary so the criterion is global. We will also adapt a
different methodology. If the velocity profile is unstable
($c_i\neq0$), then the equations with the hypothesis of $c_i=0$
should result in contradictions in some cases. Following this, a
sufficient condition for instability can be obtained.

Firstly, substituting Eq.(\ref{Eq:stable_parallelflow_Poincare})
into
Eq.(\ref{Eq:stable_stratifiedflow_TaylorGoldsteinEq_Int_Rea}), we
have
\begin{equation}
\displaystyle c_i^2\int_{a}^{b} \frac{ g(y)}{|U-c|^2}|\phi|^2\, dy
=-\int_{a}^{b}\frac{ h(y)}{|U-c|^2}|\phi|^2\,
dy,
\label{Eq:stable_stratifiedflow_Sun_Int_Rea}
 \end{equation}
where
\begin{subeqnarray}
 \displaystyle
g(y)&=&\mu+k^2+\frac{2N^2}{|U-c|^2}, \,\, and\\
\displaystyle h(y)&=&(\mu+k^2)(U-c_r)^2+U''(U-c_r)-N^2.
\label{Eq:stable_stratifiedflow_hygy}
 \end{subeqnarray}
It is noted that $g(y)>0$ for $N^2\geq0$. Then $c_i^2>0$ if
$h(y)\leq0$ throughout the domain $a\leq y \leq b$ for a proper
$c_r$ and $k$. Obviously, $h(y)$ is a monotone function of $k$:
the smaller $k$ is, the smaller $h(y)$ is. When $k=0$, $h(y)$ has
the smallest value.

\begin{equation}
h(y)=\displaystyle N^2[
\frac{(U-c_r)^2}{N^2/\mu}+\frac{U''(U-c_r)}{N^2}-1].
\label{Eq:stable_stratifiedflow_hyFroude}
\end{equation}
If we define shear, parallel and Rossby Froude numbers $Fr_t$,
$Fr_s$ and $Fr_r$ as
\begin{equation}
Fr_t^2=Fr_s^2+Fr_r^2, Fr_s^2=\displaystyle
\frac{(U-c_r)^2}{N^2/\mu}, Fr_r^2=\frac{U''(U-c_r)}{N^2},
\label{Eq:stable_stratifiedflow_Froude}
\end{equation}
where the shear  Froude number $Fr_s$ is a dimensionless ratio of
kinetic energy to potential energy. As $U''$ plays the same role
of $\beta$ effect in the Rossby wave \cite[]{SunL2006c,
SunL2007ejp}, the shear  Froude number $Fr_r$ is a dimensionless
ratio of Rossby wave kinetic energy to potential energy. Then
$h(y)\leq 0$ equals to $Fr_t^2\leq 1$. Thus a general theorem for
instability can be obtained from the above notations.

Theorem 1: If velocity $U$ and stable stratification $N^2$ satisfy
$h(y)\leq0$ or $Fr_t^2\leq 1$ throughout the domain for a certain
$c_r$, the flow is unstable with a $c_i>0$.

Physically, $Fr_t^2 \leq 1$ implies that the total kinetic energy
is smaller than the total potential energy. So the potential
energy might transfer to the kinetic energy after being disturbed,
and the flow becomes unstable. On the other hand, when the
potential energy is smaller than the kinetic energy ($Fr_t^2 >
1$), the flow is stable because no potential energy could transfer
to the kinetic energy.

Mathematically, we need derive some useful formula for
applications, since there is still unknown $c_r$ in above
equations. To this aim, we rewrite
Eq.(\ref{Eq:stable_stratifiedflow_hyFroude}) as
\begin{equation}
h(y)=\displaystyle
\mu(U+\frac{U''}{2\mu}-c_r)^2-(N^2+\frac{U''^2}{4\mu}).
\label{Eq:stable_stratifiedflow_hy}
\end{equation}
Assume that the minimum and maximum value of $U+\frac{U''}{2\mu}$
within $a\leq y\leq b$ is respectively $m_i$ and $m_a$. It is from
Eq.(\ref{Eq:stable_stratifiedflow_hy}) that $m_i\leq c_r\leq m_a$
for the smallest value of $h(y)$. Thus a general theorem for
instability can be obtained from the above notations.

Theorem 2: If velocity $U$ and stable stratification $N^2$ satisfy
$h(y)\leq0$ throughout the domain for a certain $m_i\leq c_r\leq
m_a$, there must be a $c_i>0$ and the flow is unstable.

It is from Eq.(\ref{Eq:stable_stratifiedflow_hy}) that $h(y)\leq0$
requires $\mu(U+\frac{U''}{2\mu}-c_r)^2$ less than
$(N^2+\frac{U''^2}{4\mu})$. The bigger $N^2$ is, the smaller
$h(y)$ is. So the stable stratification has a destabilization
mechanism in shear flow. This conclusion is new as former
theoretic studies always took the static stable stratification as
the stable effects for shear flows.

According to Eq.(\ref{Eq:stable_stratifiedflow_hy}), the bigger
$m_a-m_i$ is, the more stable the flow is. It is obvious that
$m_a-m_i$ is bigger for $U'/U'''>0$ than that for $U'/U'''<0$. So
the flow is more stable with the velocity profile $U'/U'''>0$.

Although Theorem 1 gives a sufficient unstable condition for
instability, the complicated expression makes it difficult for
application. In the following section we will derive simple and
useful criteria.

\section{Criteria For Flow Instability}
\begin{figure}
\centerline{
  \includegraphics[width=6cm]{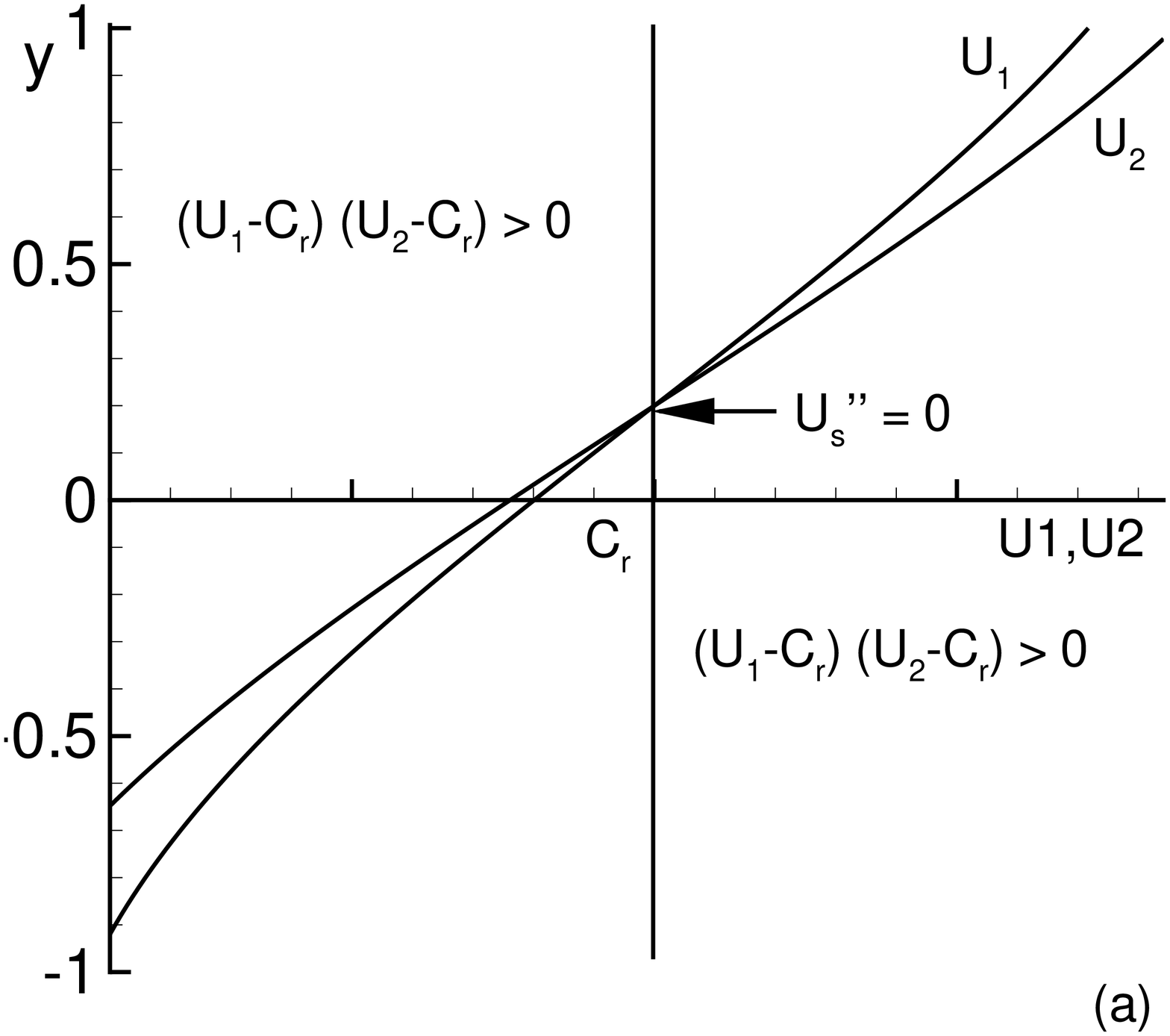}
  \includegraphics[width=6cm]{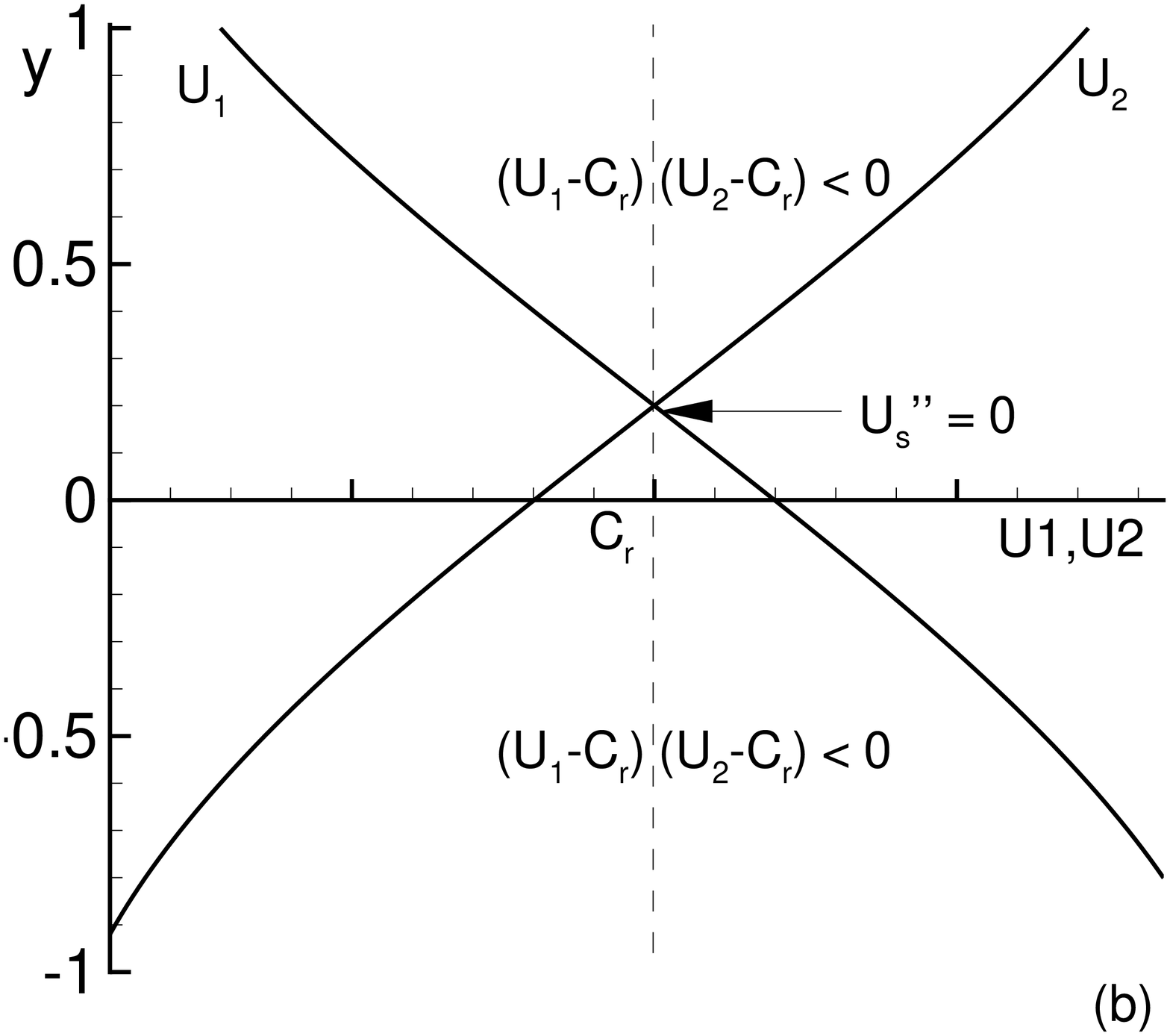}}
\caption{The value of h(y) under the condition $U''_s=0$: (a) for
$U''/(U-U_s)>0$, (b) for $U''/(U-U_s)<0$. } \label{Fig:hy}
\end{figure}

\subsection{Inviscid Flow}
The simplest flow is the inviscid shear flow with $N^2=0$. The
sufficient condition for instability is $h(y)\leq0$. To find such
condition, we rewrite $h(y)$ in
Eq.(\ref{Eq:stable_stratifiedflow_hy}) as
\begin{equation}
h(y)=(U_1-c_r)(U_2-c_r) \label{Eq:Rayleigh-hy}
 \end{equation}
where $U_1=U$ and $U_2=U+U''/\mu$. Then there may be three cases.
Two of them have $U_1$ intersecting with $U_2$ at $U''_s=0$
(Fig.\ref{Fig:hy}). The first case is that $U''/(U-U_s)>0$; thus,
$h(y)>0$ always holds at $c_r=U_s$ as shown in Fig.\ref{Fig:hy}a.
The second case is that $U''/(U-U_s)<0$; thus, $h(y)<0$ always
holds in the whole domain, as shown in Fig.\ref{Fig:hy}b. In this
case, the flow  might be unstable.

The sufficient condition for instability can be found from
Eq.(\ref{Eq:Rayleigh-hy}) as shown in Fig.\ref{Fig:hy}b. Given
$c_r=U_s$, Eq.(\ref{Eq:Rayleigh-hy}) becomes
\begin{equation}
h(y)=\frac{(U-U_s)^2}{\mu}[\mu+\frac{U''}{(U-U_s)}]
 \end{equation}
If $\frac{U''}{(U-U_s)}<-\mu$ is always satisfied, $h(y)<0$ holds
within the domain.

Corollary 1.1: If the velocity profile satisfies
$\frac{U''}{U-U_s}<-\mu$ within the domain, the flow is unstable.

Since \cite{SunL2007ejp} obtained a sufficient condition for
stability, i.e. $\frac{U''}{U-U_s}>-\mu$ within the domain. The
above condition for instability is nearly marginal
\cite[]{SunL2008cpl}.

The last case is that $U''\neq 0$ throughout the domain; thus,
$h(y)>0$  always exists somewhere within the domain, as shown in
Fig.\ref{Fig:TG-hy}a.

\subsection{Stably Stratified Flow}
\begin{figure}
  \centerline{\includegraphics[width=6cm]{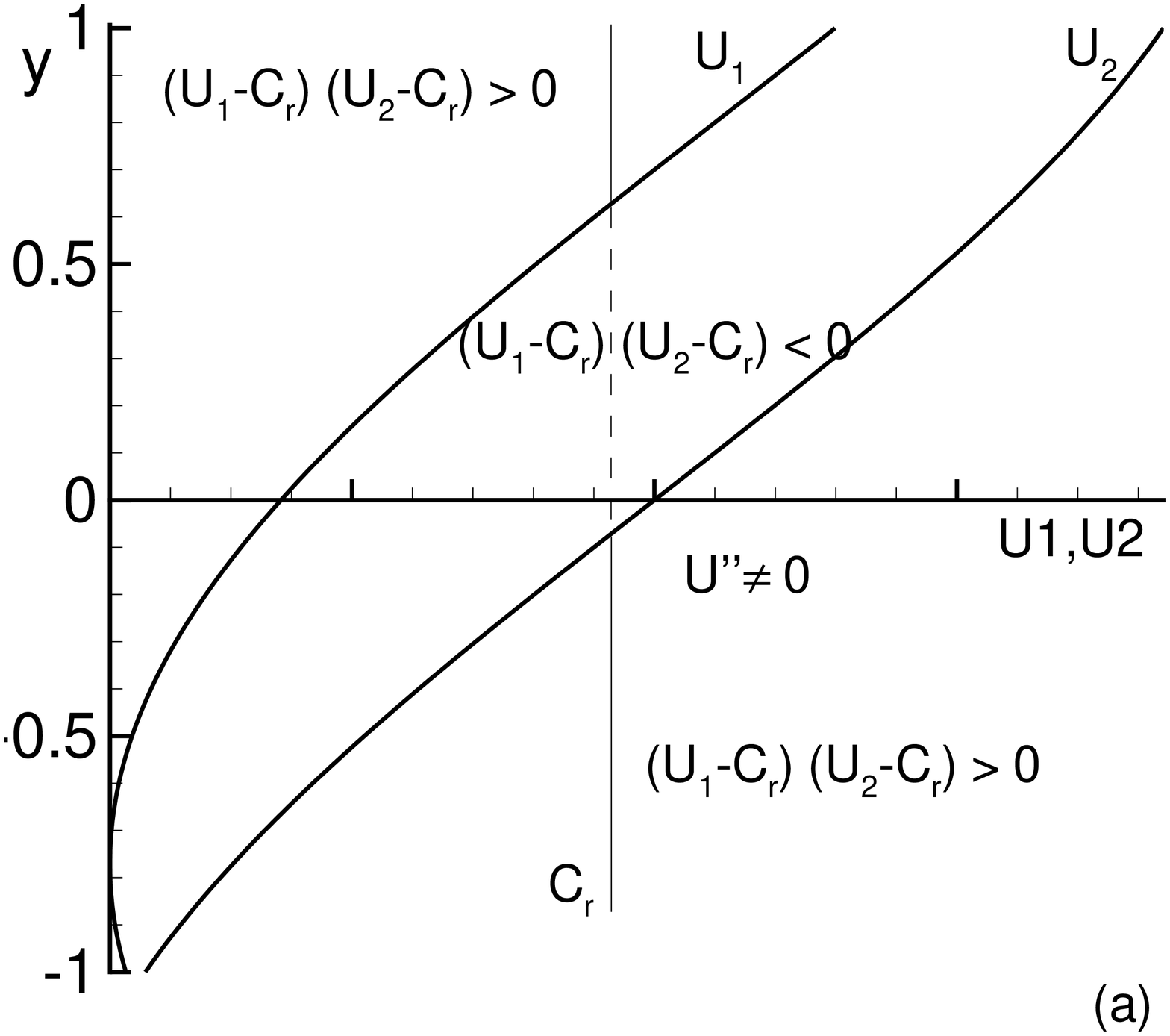}
  \includegraphics[width=6cm]{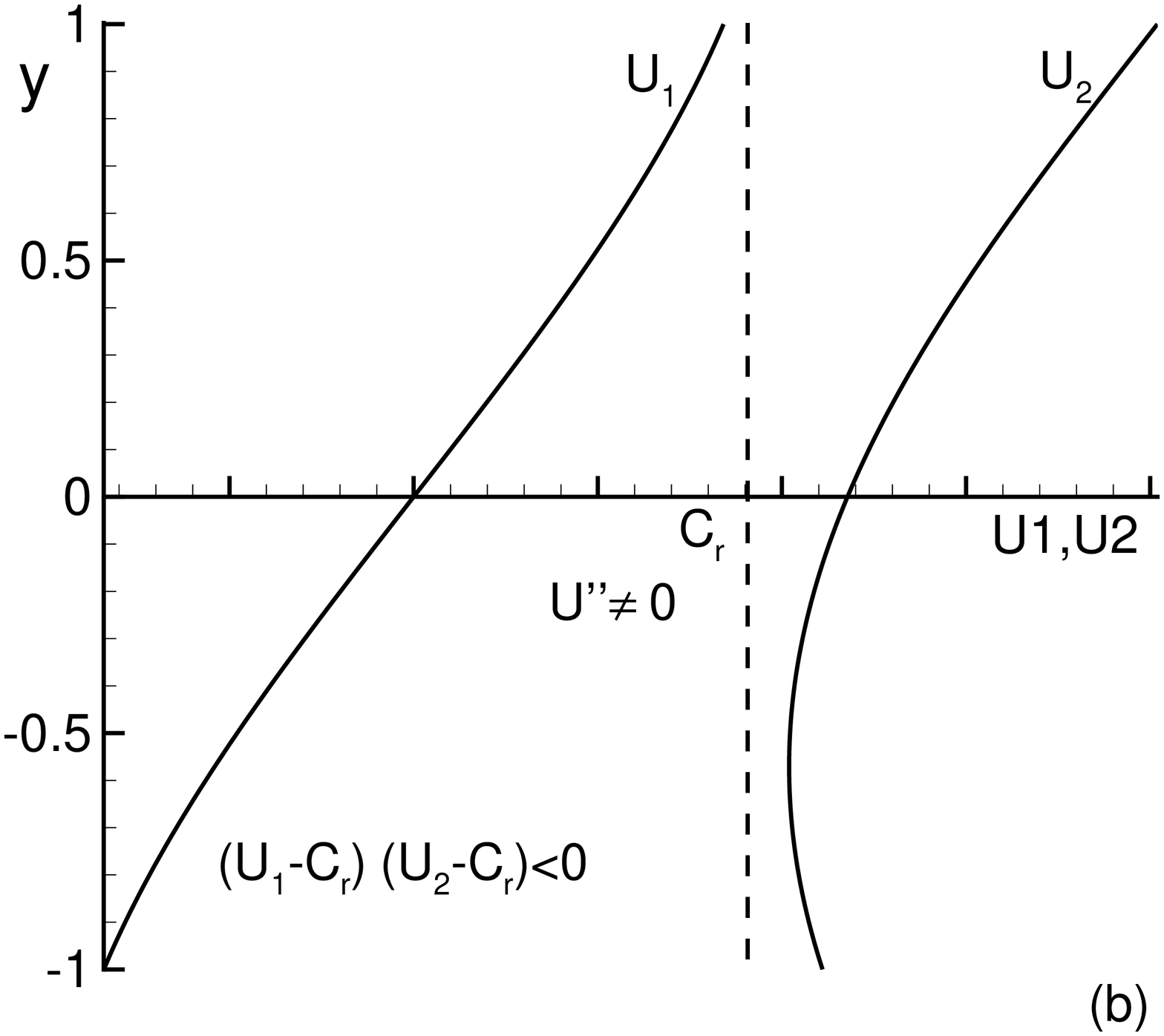}}
\caption{The value of $h(y)$ for $U''\neq0$ in case 3 and case 4.}
\label{Fig:TG-hy}
\end{figure}

If the static stratification is stable ($N^2>0$), then $g(y)$ is
positive. The flow is unstable if $h(y)$ is negatively defined
within $a\leq y \leq b$ at $k=0$. We rewrite $h(y)$ as
\begin{equation}
\begin{array}{rl}
 h(y)=&\mu (U_1-c_r)(U_2-c_r) \\
    =&\displaystyle \mu [U+\frac{1}{2\mu}(U''-\sqrt{U''^2+4\mu N^2}\,)-c_r]\\
    &\displaystyle \times
[U+\frac{1}{2\mu}(U''+\sqrt{U''^2+4\mu N^2}\,)-c_r].
\end{array}\label{Eq:TaylorGoldsteinEq-hy}
 \end{equation}
The value of $h(y)$ can be classified into 4 cases. The first and
the second ones ($U''_s=0$ and $N^2_s=0$ at $y=y_s$) are similar
to discussed above and shown in Fig\ref{Fig:hy}a and
Fig\ref{Fig:hy}b. For such cases, we have a sufficient condition
for instability,
\begin{equation}
\frac{U''(U-U_s)-
N^2}{(U-U_s)^2}<-\mu.\label{Eq:TaylorGoldsteinEq-SIC1}
 \end{equation}
This can be derived directly from
Eq.(\ref{Eq:stable_stratifiedflow_Sun_Int_Rea}), similar to
Corollary 1.1.  The first sufficient condition for instability is
due to the shear instability, and the unstable criterion is
Eq.(\ref{Eq:TaylorGoldsteinEq-SIC1}).

Corollary 1.2: If the velocity profile satisfies
$\frac{U''(U-U_s)-N^2}{(U-U_s)^2}<-\mu$ within the domain, the
flow is unstable.

The third case ($U''\neq 0$) is also similar to the case in
Fig\ref{Fig:TG-hy}a, and the flow is stable. The last one is
unstable flow shown in Fig.\ref{Fig:TG-hy}b, where $U''\neq 0$ and
$h(y)<0$ throughout. In the last case, the maximum of $U_1$ must
be smaller than the minimum of $U_2$ so that a proper $c_r$ within
the $U_1$ and $U_2$ could be used for the unstable waves. Although
the exact criterion can not be obtained as the required maximum
and minimum can not be explicitly given, the approach is very
straightforward.

Nevertheless, we can also obtain some approximate criterion for
the fourth case. It is from Eq.(\ref{Eq:stable_stratifiedflow_hy})
that $h(y)\leq0$ if the minimax of $\mu(U+\frac{U''}{2\mu}-c_r)^2$
is less than the minimum of $(N^2+\frac{U''^2}{4\mu})$. As the
minimax value of $\mu(U+\frac{U''}{2\mu}-c_r)^2$ is
$\frac{1}{4}\mu(m_a-m_i)^2$ when $c_r=(m_a+m_i)/2$, we obtained a
new criterion according to
Eq.(\ref{Eq:stable_stratifiedflow_Froude}).
\begin{equation}
Fr_t^2(c_r) = \frac{1}{4\mu N^2}[\mu^2 (m_a+m_i)^2-U''^2].
\label{Eq:stable_stratifiedflow-Fs2}
 \end{equation}
Thus a sufficient (but not necessary) condition for $h(y)<0$ is
that the following equation holds for $a\leq y\leq b$.
\begin{equation}
Fr_t^2\leq 1. \label{Eq:stable_stratifiedflow-Criterion}
\end{equation}

From the above corollaries, the flow might be unstable if the
static stable stratification is strong enough. The stably
stratification destabilize the flow, which is a new unstable
mechanism. The above corollary contradicts the previous results
\cite[]{Abarbanelt1984}, but it agrees well with the recent theory
\cite[]{Friedlander2001}, experiments \cite[]{Zilitinkevich2008}
and simulations \cite[]{Alexakis2009}. Again, we point out here
that the flow is unstable due to potential energy transfer to
kinetic energy under the condition of $Fr_t^2 \leq 1$.

This conclusion is new because it is quite different from previous
theorems in which the static stable stratification plays the role
as a stabilizing factor for shear flows.

\section{Discussion}

\subsection{Necessary Instability Criterion}
In the above investigation, it was found that stable
stratification is a destabilization mechanism for the flow. Such
finding is not surprising if one notes the terms in
Eq.(\ref{Eq:stable_stratifiedflow_TaylorGoldsteinEq}).
Mathematically the sum of terms in square brackets should be
negative for the wave solution. Thus both $\frac{U''}{U-c}>0$ and
$N^2<0$ do favor this condition. This is why the unstable
solutions always occur at $\frac{U''}{U-c}<0$ in shear flow. And
$N^2>0$ here might lead to $c_i^2>0$. Physically, the perturbation
waves are truncated in the neutral stratified flow. But the stable
stratification allows wide range of waves in the perturbation.
Such waves might interact with each other like what was
illustrated in \cite{SunL2008cpl}.

As Theorem 1 is the only sufficient condition, it is hypothesized
that the criterion is not only the sufficient but also the
necessary condition for instability in stably stratified flow.
This hypothesis might be criticized in that the flow might be
unstable ($c_i^2>0$) if $h(y)$ changes sign within the interval
(Fig\ref{Fig:TG-hy}a), where a proper chosen $\phi$ would let the
right hand of Eq.(\ref{Eq:stable_stratifiedflow_Sun_Int_Rea})
become negative.

However, this criticism is not valid for the case in
Fig\ref{Fig:TG-hy}a. It is from the well-known criteria (e.g.
Rayleigh's inflexion point theorem) that the proper chosen $\phi$
always let the right hand of
Eq.(\ref{Eq:stable_stratifiedflow_Sun_Int_Rea}) vanish. It seems
that the flow tends to be stable, or the perturbations have a
prior policy to let $c_i=0$. The flow become unstable unless any
choice of $\phi$ would let the right hand of
Eq.(\ref{Eq:stable_stratifiedflow_Sun_Int_Rea}) be negative. In
this situation, we hypothesize that Theorem 1 fully solves the
stability problem.

\subsection{Long-wave Instability}
In inviscid shear flows, it has been recognized that very
short-wave perturbations are dynamically stable under neutral
stratification, and the dynamic instability is due to the larger
wavelengths \cite[]{SunL2006b}. It should be noted that Rayleigh's
case is reduced to the Kelvein-Helmholtz vortex sheet model under
the long-wave limit $k\ll 1$
\cite[]{Huerre1998,CriminaleBook2003}. We have shown that this can
be extended to shear flows, and that the growth rate $\omega_i$,
is proportional to $\sqrt{\mu_1}$ \cite[]{SunL2006b, SunL2008cpl}.

Such conclusion can be simply generated to the stratified shear
flows, which can be seen from
Eq.(\ref{Eq:stable_stratifiedflow_hygy}). If $k$ is larger than a
critical value $k_c$, the sufficient condition in Theorem 1 can
not be satisfied and the flow is stable. For shortwave ($k\gg1$),
$h(y)$ is always larger than that for long-wave $k\ll 1$. The
long-wave instability in the stratified shear flow was also noted
by \cite{Miles1961,Miles1963} and \cite{Howard1961}, who showed a
likelihood of $c_i\rightarrow 0$ at $k\rightarrow\infty$. The
long-wave instability theory can explain the results in numerical
simulations \cite[]{Alexakis2009}, where the unstable
perturbations are long-wave.

\subsection{Local Criterion}
In the above investigations, an parameter $\mu$ is used, which
represents the ratio of two integrations with boundaries. So the
criteria are global. On the other hand, we can also investigate
the local balance without boundary conditions. For example,
consider the flow within a thick layer $-\delta \leq y \leq
\delta$. The velocity is $U(y)=U_0+U'y$, and the kinetic energy is
$(U-c_r)^2$. The stratification is $N^2$, and the potential energy
is $N^2 d^2$, where $d=2\delta$ is the thickness of the layer. The
Froude number is $Fr_t^2=(U'^2\delta^2)/(N^2d^2)$ for $c_r=U_0$.
The instability criterion in
Eq.(\ref{Eq:stable_stratifiedflow-Criterion}) becomes

\begin{equation}
Ri=\frac{U'^2}{N^2}>\frac{1}{4}.
\label{Eq:stable_stratifiedflow-Criterion-Ri}
\end{equation}
If local gradient Richardson number exceeds $1/4$, the local
disturbances is unstable. However, the flow might be stable as the
globe total  Froude number $Fr_t^2>1$. This criterion is opposite
to Miles-Howard theorem, we will show why Miles-Howard' theorem is
not correct from their derivations.

\subsection{Relations to Other Theories}
In the inviscid shear flow, the linear theories, e.g.,
Rayleigh-Kuo cirterion \cite[]{CriminaleBook2003},  Fj\o rtoft
criterion \cite[]{Fjortoft1950} and Sun's criterion
\cite[]{SunL2007ejp}, are equal to Arnol'd's nonlinear stability
criterion \cite[]{Arnold1965a}. Arnol'd's first stability theorem
corresponds to Fj\o rtoft's criterion
\cite[]{Drazin2004,CriminaleBook2003}, and Arnol'd's second
nonlinear theorem corresponds to Sun's criterion
\cite[]{SunL2007ejp,SunL2008cpl}. It is obvious that the present
theory, especially Corollary 1.1 is a natural generalization of
inviscid theories.

In the stratified flow, \cite{Miles1961,Miles1963} and
\cite{Howard1961} applied a transform $F=\phi/(U-c)^n$ to
Eq.(\ref{Eq:stable_stratifiedflow_TaylorGoldsteinEq}), which
allows different kind of perturbations. Thus $n=1/2$ gives Miles's
theory and $n=1$ gives Howard's semicircle theorem.

Considering that $n=1$ and $N^2=0$ \cite[]{Howard1961},
Eq.(\ref{Eq:stable_stratifiedflow_TaylorGoldsteinEq_Int_Rea})
 becomes
\begin{equation}
\displaystyle\int_{a}^{b} (|F'|^2+k^2|F|^2)[(U-c_r)^2-c_i^2] dy=0.
\label{Eq:stable_stratifiedflow_TaylorGoldsteinEq_Miles_Rea}
 \end{equation}
It is from
Eq.(\ref{Eq:stable_stratifiedflow_TaylorGoldsteinEq_Miles_Rea})
that all the inviscid flows (no mater what the velocity profile
$U(y)$ is) must be temporal unstable ($k$ is real). This
contradicts the criteria (both linear and nonlinear ones) for
inviscid shear flow. So the wavenumber $k$ in
Eq.(\ref{Eq:stable_stratifiedflow_TaylorGoldsteinEq_Miles_Rea})
should be complex $k=k_r+ik_i$. Besides, from
Eq.(\ref{Eq:stable_stratifiedflow_hy}),
Eq.(\ref{Eq:TaylorGoldsteinEq-hy}) and Fig.\ref{Fig:TG-hy}b, the
unstable $c_r$ might be either within the value of $U$ or beyond
the value of $U$. This also contradicts the Howard's semicircle
theorem for the stratified flow. It implies that the transform $F$
is not suitable for temporal stability problem.

Taking $n=1/2$, Howard extracted a new equation from
Taylor-Goldstein equation,
\begin{equation}
\displaystyle
[(U-c)F']'-[k^2(U-c)+\frac{U''}{2}+(\frac{1}{4}U'^2-N^2)/(U-c)]F=0
\label{Eq:stable_stratifiedflow_TaylorGoldsteinEq_Howard}
 \end{equation}
After multiplying above equation by the complex conjugate of $F$
and integrating over the flow regime, then the imaginary part of
the expression is
\begin{equation}
\displaystyle -c_i \int_a^b
|F'|^2+[k^2|F|^2+(\frac{1}{4}U'^2-N^2)|F|^2/|U-c|^2=0
\label{Eq:stable_stratifiedflow_TaylorGoldsteinEq_Howard_Img}
 \end{equation}
Miles-Howard theorem concludes that if $c_i \neq 0$, then
$Ri<\frac{1}{4}$ for instability.

However, the transform $F=\phi/\sqrt{U-c}$ requires a complex
function $F$, even though both $\phi$ and $c$ are real. In that
$\sqrt{U-c}$ might be complex somewhere as $U-c_r<0$.
Consequently, the wave number $k$ in
Eq.(\ref{Eq:stable_stratifiedflow_TaylorGoldsteinEq_Howard}) is a
complex number but no longer a real number as that assumed in
Taylor-Goldstein equation. The complex wavenumber $k$ leads to
spatial stability problem but temporal stability problem
investigated in this study. The assumption of $c_i=0$ with $k_i
\neq0$ implies the flow is unstable with $\omega_i \neq0$.
However, Howard ignored this in his derivations. That is why
Miles-Howard theorem leads contradictions to the present studies.

Although the transform $F=\phi/(U-c)^n$ leads some contradictions
with Rayleigh criterion and present results, it might be useful
for the viscous flows. In these flows, the spatial but temporal
stability problem is dominated, and $k=k_r+ik_i$ is the complex
wavenumber. It is well known that the plane Couette flow is
viscously unstable for Reynolds number $Re> Re_c$ from the
experiments but viscously stable from the Orr-Sommerfeld equation
\cite[]{CriminaleBook2003}.  If applying the transform in
Eq.(\ref{Eq:stable_stratifiedflow_TaylorGoldsteinEq_Miles_Rea})
all the inviscid flows must be unstable. Thus the plane Couette
flow might be stable only for $Re< Re_c$ due to the stabilization
of the viscosity.

It is argued that the Taylor-Goldstein equation represents
temporal instability, the transform represents spatial instability
\cite[]{Huerre1998,CriminaleBook2003}. In that the perturbation is
seen along with the flow at the speed of $(U-c)$ in
\cite{Miles1961,Howard1961}. The transform $F=\phi/(U-c)^n$ also
turns real wavenumber $k$ into complex number, $c_i=0$ implies
$\omega_i \neq 0$. The assumption of real $k$ after transform will
leads to contradictions with the results derived from the
Taylor-Goldstein equation. So the previous investigators can
hardly generalize their results from homogeneous fluids to
stratified fluids.

\section{Conclusion}
In summary, the stably stratification is a destabilization
mechanism, and the flow instability is due to the competition of
the kinetic energy with the potential energy. Globally, the flow
is always unstable when the total Froude number $Fr_t^2\leq 1$,
where the larger potential energy might transfer to the kinetic
energy after being disturbed. Locally, the flow is unstable as the
gradient Richardson number $Ri>1/4$. The approach is very
straightforward and can be used for similar analysis. In the
inviscid stratified flow, the unstable perturbation must be
long-wave scale. This result extends the Rayleigh's,
Fj{\o}rtoft's, Sun's and Arnol'd's criteria for the inviscid
homogenous fluid, but contradicts  the well-known Miles and Howard
theorems. It is argued here that the transform $F=\phi/(U-c)^n$ is
not suitable for temporal stability problem, and that it will
leads to contradictions with the results derived from
Taylor-Goldstein equation.

The author thanks Dr. Yue P-T at Virginia Tech, Prof. Yin X-Y at
USTC, Prof. Wang W. at OUC and Prof. Huang R-X at WHOI for their
encouragements. This work is supported by the National Basic
Research Program of China (No. 2012CB417402), and the Knowledge
Innovation Program of the Chinese Academy of Sciences (No.
KZCX2-YW-QN514).

\bibliographystyle{jfm}

\end{document}